# Elastic, electronic and optical properties of hypothetical SnNNi$_3$ and CuNNi$_3$ in comparison with superconducting ZnNNi$_3$

## M.A. Helal, A.K.M.A. Islam[*]

*Department of Physics, Rajshahi University, Rajshahi 6205, Bangladesh*

**A B S T R A C T**

The elastic, electronic and optical properties of MNNi$_3$ (M= Zn, Sn and Cu) have been calculated using the plane-wave ultrasoft pseudopotential technique which is based on the first-principles density functional theory (DFT) with generalized gradient approximation (GGA). The optimized lattice parameters, independent elastic constants ($C_{11}$, $C_{12}$, and $C_{44}$), bulk modulus $B$, Compressibility $K$, shear modulus $G$, and Poisson's ratio $v$, as well as the band structures, total and atom projected densities of states and finally the optical properties of MNNi$_3$ have been evaluated and discussed. The electronic band structures of the two hypothetical compounds show metallic behavior just like the superconducting ZnNNi$_3$. Using band structures, the origin of features that appear in different optical properties of all the three compounds have been discussed. The large reflectivity of the predicted compounds in the low energy region might be good candidate materials as a coating to avoid solar heating.

## 1. Introduction

The unexpected appearance of superconductivity with $T_c \sim$ 8K for Ni-rich ternary carbide MgCNi$_3$ [1] with cubic antiperovskite-like structure has initiated a lot of interest among the scientific community [2-36]. The investigations of a set of ternary carbon containing anti-perovskites MCNi$_3$ have been performed, not only due to existence of Ni-rich element but also due to search of the origin of exact nature of the superconducting state. All these attempts ultimately led to successful synthesis and some theoretical examinations of a set of Ni-rich carbides such as ZnCNi$_3$, CdCNi$_3$, MgCNi$_3$ etc. [11-24]. A detailed study of the properties of MgCNi$_3$ and related antiperovskite-type carbides [2-6] and efforts to improve their properties through chemical substitution [7-10] have also been carried out.

Recently, a new type of Ni-rich carbon free superconductor ZnNNi$_3$ with $T_c$ = 3K [25] has been synthesized, which is the only superconducting Ni-based nitrogen containing material. This discovery is intriguing in view of the rigid-band picture [32]. Here one can consider the antiperovskite ZnNNi$_3$ as a one electron doped superconducting MCNi$_3$ phase, where the Fermi level should be located far from the Ni-3$d$ peak, i.e., in the region of a quite low DOS, which is unfavorable for superconductivity. In addition, as far as we know, ZnNNi$_3$ is the first nitrogen containing superconducting material in the Ni-based anti-perovskite series. The discovery of superconductivity has given strong motivation to study the Ni-based antiperovskite series.

Further, very recently a set of antiperovskite-like Ni-rich ternary nitrides MNNi$_3$ (M = Zn, Mg, Cd, Al, Ga and In) has been synthesized and characterized by means of band structure calculations [25-35]. Also the mechanical properties of twelve MNNi$_3$-type compounds with M = Zn, Mg, Cd, Al, Ga , In, Sn, Sb, Pd, Cu, Ag and Pt have been examined theoretically [37]. But no work has been done on electronic and optical properties of several compounds e.g. CuNNi$_3$ and SnNNi$_3$, for example. Despite several works on elastic and some electronic properties of the superconducting ZnNNi$_3$, no attempt has been made to calculate its optical properties.

In view of these circumstances, we perform a first-principles study to predict the elastic, electronic and optical properties of MNNi$_3$ (M = Zn, Cu and Sn). The calculated properties of the two hypothetical SnNNi$_3$ and CuNNi$_3$ are then compared with those of the superconducting ZnNNi$_3$.

## 2. Computational methods

All the calculations are performed using the first-principles pseudopotential method in the framework of density functional theory (DFT) with generalized gradient approximation as implemented in the CASTEP code [41]. The basis set of valence electronic states was taken to be $3d^84s^2$ for Ni, $2s^22p^3$ for N, and $3d^{10}4s^2$, $5s^25p^2$, and $3d^{10}4s^1$ for Zn, Sn, and Cu, respectively. The elastic constants are calculated by the 'stress-strain' method. All the calculating properties for MNNi$_3$ (M = Zn, Sn, and Cu) used a plane-wave cutoff energy 500 eV and 15×15×15 Monkhorst-Pack [42] grid for the sampling of the Brillouin zone. Geometry optimization is conducted using convergence thresholds of $5\times10^{-6}$ eV/atom for the total energy, 0.01 eVÅ$^{-1}$ for the maximum force, 0.02 GPa for maximum stress and $5\times10^{-4}$ Å for maximum displacement.

The Ni-based ZnNNi$_3$ superconductor adopts a cubic structure (space group *Pm-3m*) consisting Zn atoms at the corners, N at the body center and Ni at the face centers of the cube. The atomic positions are Zn: 1*a* (0,0,0); N: 1*b* (1/2,1/2,1/2); Ni: 3*c* (1/2,1/2,0) (Fig. 1). The hypothetical ternary nitrides MNNi$_3$ with M = Sn and Cu like other synthesized similar compounds are examined in the same cubic antiperovskite structure.

## 3. Results and discussion

### 3.1. Lattice constants and elastic properties

The geometry optimization was carried out as a function of the normal stress by minimizing the total energy of the two hypothetical compounds along with the ZnNNi$_3$. The procedures lead to successful optimization of the two hypothetical structures. The optimized parameters and elastic constants for all the three antiperovskite-type compounds MNNi$_3$ (M = Zn, Sn and Cu) are shown in Table 1. For ZnNNi$_3$, our calculated value of the lattice constant (3.784 Å) is in good agreement with experiment, [25]. It is seen that a relationship among the lattice constants for the three compounds are as follows:

$$a(\text{CuNNi}_3) < a(\text{ZnNNi}_3) < a(\text{SnNNi}_3).$$

**Table 1.** The calculated lattice constants (*a*, in Å), volume (*V*, in Å$^3$), independent elastic constants ($C_{ij}$, in GPa), bulk modulus (*B*, in GPa), compressibility (*K*, in GPa$^{-1}$), shear modulus (*G*, in GPa) and Poisson's ratio (υ) for synthesized and hypothetical MNNi$_3$, where M are Zn, Sn and Cu.

| Parameters | ZnNNi$_3$ | SnNNi$_3$ | CuNNi$_3$ |
|---|---|---|---|
| *a* | 3.784, 3.756[1] | 3.927, 3.910[2] | 3.761, 3.745[2] |
| *V* | 54.19 | 60.58 | 53.18 |
| $C_{11}$ | 335.62, 354.3[3] | 262.14, 239.9[2] | 349.23, 396.9[2] |
| $C_{12}$ | 124.84, 134.0[3] | 134.24, 153.5[2] | 128.14, 121.9[2] |
| $C_{44}$ | 48.43, 48.1[3] | 35.47, 2.2[2] | 32.58, 7.8[2] |
| *B* | 195.1, 204.9[2] | 176.8, 182.3[2] | 201.8, 213.6[2] |
| *K* | 0.0051, 0.0049[2] | 0.0056, 0.0055[2] | 0.0049, 0.0047[2] |
| *G* | 48.43 | 35.47 | 32.58 |
| υ | 0.346 | 0.381 | 0.376 |

[1] Expt. [25], [2] WIEN2K code [37], [3] VASP code [31].

The values of three independent elastic constants ($C_{11}$, $C_{12}$, and $C_{44}$) for cubic lattice were calculated by applying a proper strain to the equilibrium structure from different directions [32-35]. These values are included in Table 1. All these values are positive and satisfy the generalized criteria [38] for mechanically stable crystals: $(C_{11}-C_{12}) > 0$; $(C_{11} + 2C_{12}) > 0$; $C_{44} > 0$. These conditions also lead to a restriction on the value of the bulk modulus $B$, which is required to be in between $C_{11}$ and $C_{12}$, i.e., $C_{12} < B < C_{11}$. The values of the bulk modulus $B$ and shear modulus $G$ were calculated by $(C_{11} + 2C_{12})/3$ and $C_{44}$, respectively. From Table 1 we see that the value of the bulk modulus $B$ and lattice constants follows the well-known relationship $B \sim V^{-1}$ [39]. According to our calculations the maximum bulk modulus (minimum compressibility) is obtained for $CuNNi_3$, whereas, the minimum bulk modulus (maximum compressibility) is obtained for $SnNNi_3$, i.e. $B(SnNNi_3) < B(ZnNNi_3) < B(CuNNi_3)$. The shear modulus is obtained in the inverse order of the bulk modulus. The values of the Poisson's ratio ($\upsilon$) are in the interval 0.346 - 0.381. As the Poisson's ratio ($\upsilon$) for the brittle covalent materials is small, and for ductile metallic materials it is typically 0.33 [40], we see that all these compounds belong to metallic like systems.

## 3.2. Electronic properties

The calculated band structures along high symmetry directions (Γ-X-M-Γ-R-X) of the brillouin zone for Ni-rich ternary nitrides $MNNi_3$ (M = Zn, Sn, and Cu) are shown in Figs.1 (a- c). From these figures we see that at least one energy level cross the Fermi level, indicating that all the considered compounds exhibit metallic properties. The valence region for $ZnNNi_3$, which extends from -8.1 eV (not shown) up to the Fermi level $E_F = 0$ eV is composed mainly of Ni-3$d$ and N-2$p$ states. N-2$s$ states are situated in the region far from the Fermi level, whereas Zn atom has small contribution in the low energy level. However, both of these states play a relatively minor role in the valence area. But the valence region of $SnNNi_3$ and $CuNNi_3$ are composed of Ni-3$d$, N-2$p$ and Sn-5$s$ and Cu-3$d$ states. For all the cubic phases, N-2$p$ states are partially hybridized with Ni-3$d$ states in the valence region. Thus we can conclude that the band structures of these compounds are mainly dominated by Ni and N.

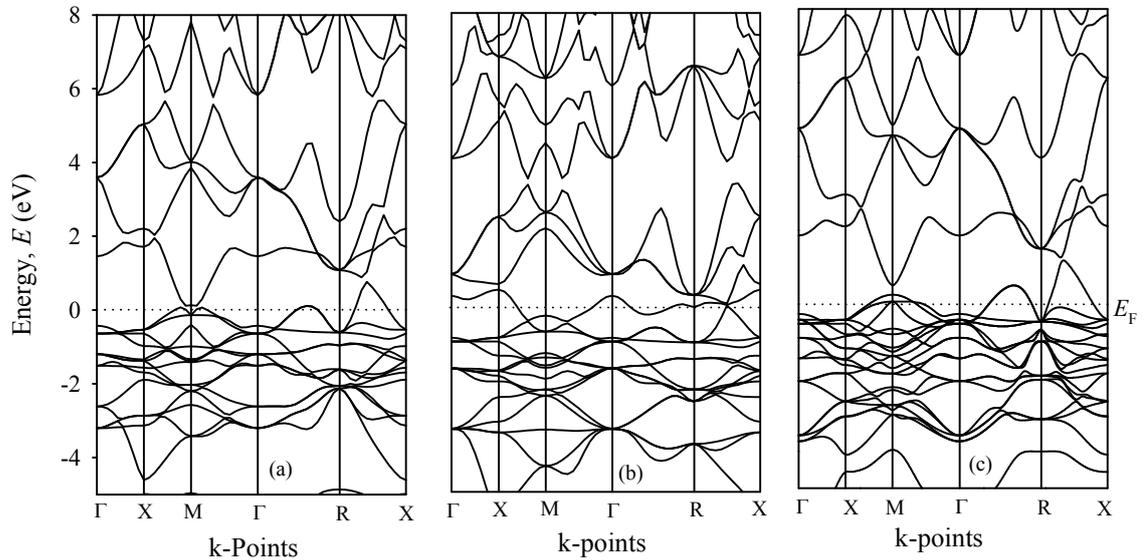

**Fig. 1.** The calculated band structures of (a) $ZnNNi_3$, (b) $SnNNi_3$ and (c) $CuNNi_3$.

In order to elucidate the different contributions from the different components in the materials to the conductivity, the DOS at the Fermi level $N(E_F)$ and atomic contributions are plotted in Figs. 2 (a-c). It is seen from these figures that Ni-3$d$ and N-2$p$ bands are strongly hybridized near the Fermi level. From the partial density of states, we see that, for $ZnNNi_3$ and $CuNNi_3$, Zn and Cu atoms have no contribution to the total DOS. But for $SnNNi_3$, it is noticeable that Sn-5$s$ has a contribution to the total DOS. So, the highest total DOS is obtained for $SnNNi_3$.

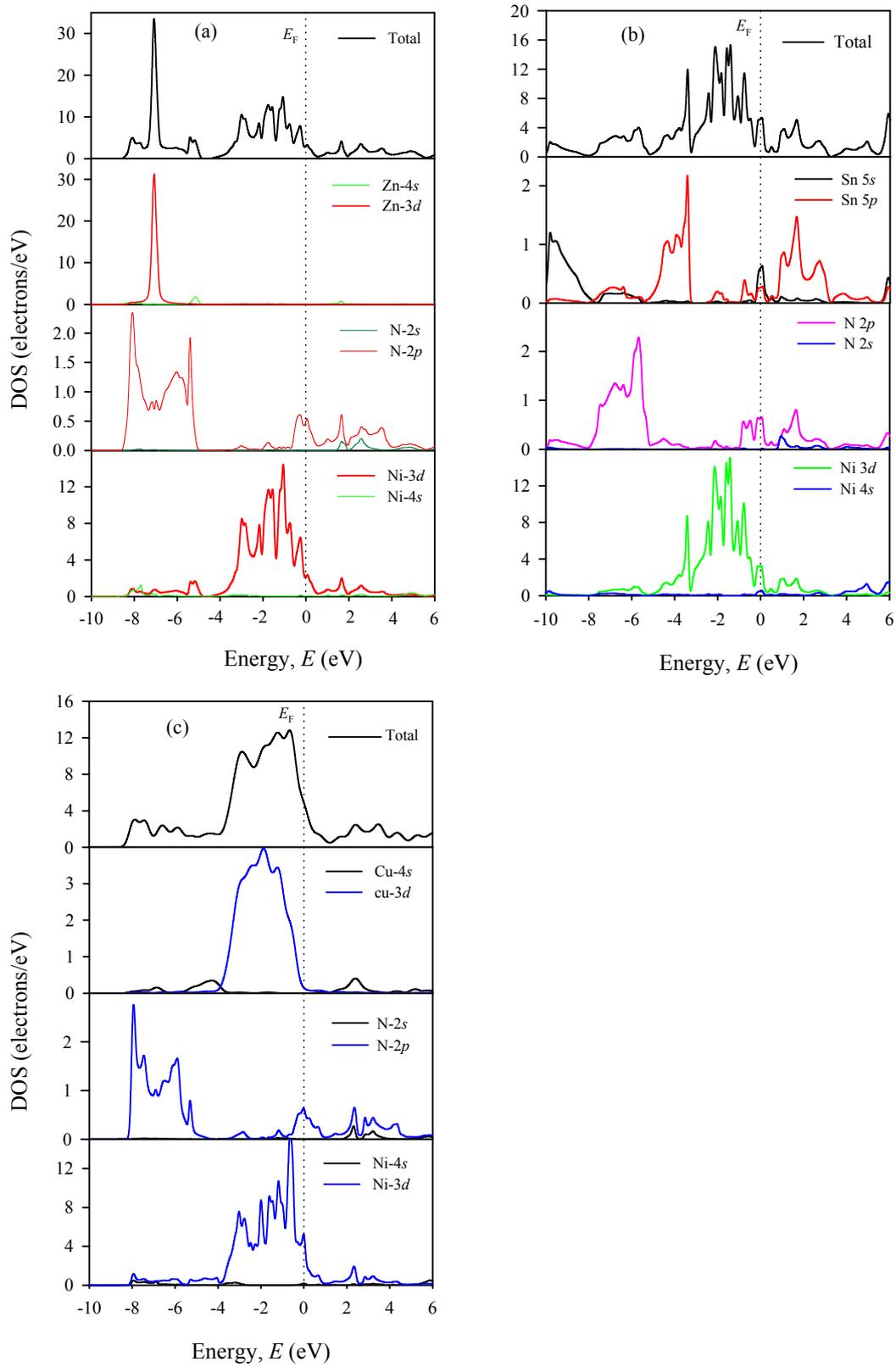

**Fig. 2.** Total and partial density of states of (a) ZnNNi$_3$, (b) SnNNi$_3$, and (c) CuNNi$_3$.

## 3.3. Optical properties

The study of the optical functions helps to give a better understanding of the electronic structure. Fig. (3-7) show the optical functions of MNNi$_3$ (M = Zn, Sn, and Cu) calculated for photon energies up to 40 eV. We have used a 0.5 eV Gaussain smearing for all calculations. This smears out the Fermi level, so that k-points will be more effective on the Fermi surface.

Fig. 3 shows the reflectivity spectra of MNNi$_3$ (M = Zn, Sn, and Cu). We notice that the reflectivity is ~ 0.65 - 0.40 in the infrared region and the value drops in the high energy region with some peaks as a result of interband transition. For energy above 13 eV, CuNNi$_3$ has relatively larger reflectivity compared to that of ZnNNi$_3$. The large reflectivity for $E < 1$ eV indicates the characteristics of high conductance in the low energy region.

For ZnNNi$_3$, the real part $\varepsilon_1(\omega)$ of the dielectric function (Fig. 4a) vanishes at about 16 eV. This corresponds to the energy at which the reflectivity (Fig. 3a) exhibits a sharp drop and the energy loss function (describing the energy loss of a fast electron traversing in the material) shows a first peak (Fig. 5a). This peak in energy-loss function at about 16 eV arises as $\varepsilon_1(\omega)$ goes through zero and $\varepsilon_2(\omega)$ is small at such energy, thus fulfilling the condition for plasma resonance at 16 eV ($\hbar\omega_p = 16$ eV). For SnNNi$_3$, it is seen that the real part $\varepsilon_1(\omega)$ of the dielectric functions (Fig. 4b) vanishes at about 17 eV. This is the energy at which the reflectivity (Fig. 3b) exhibits a sharp drop and the energy loss function (Fig. 5b) shows a first peak. Thus the plasma frequency for SnNNi$_3$ is 17 eV. Similarly it is also observed that the plasma frequency for CuNNi$_3$ is 26.5 eV. The materials become transparent when the frequency of the incident light is higher than the plasma frequency. The electron energy loss function is an important optical parameter describing the energy loss of a fast electron traversing a material. For each compound the peak in the dielectric functions around 0.5 eV is due to the transitions within the Ni-3$d$ bands. The large negative values of $\varepsilon_1(\omega)$ indicates that all the three MNNi$_3$ crystal have a Drude-like behavior.

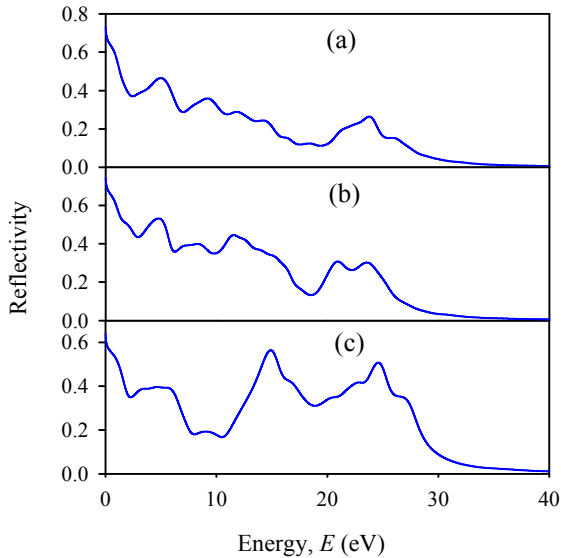
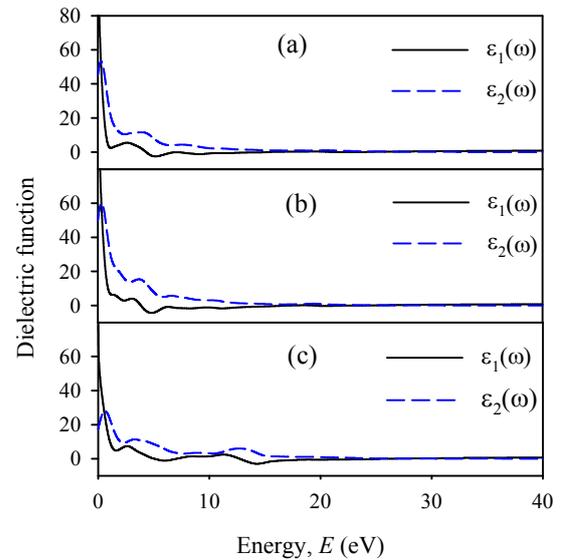

**Fig. 3.** Reflectivity spectra of (a) ZnNNi$_3$, (b) SnNNi$_3$, and (c) CuNNi$_3$.

**Fig. 4.** Dielectric functions of (a) ZnNNi$_3$, (b) SnNNi$_3$, and (c) CuNNi$_3$.

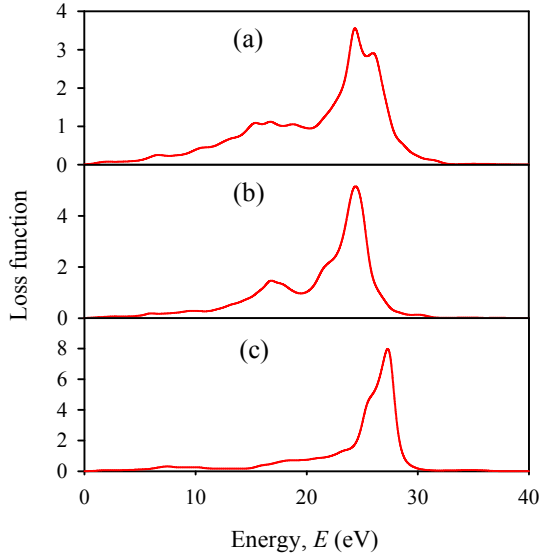

**Fig. 5.** Loss functions of (a) ZnNNi$_3$, (b) SnNNi$_3$ and (c) CuNNi$_3$.

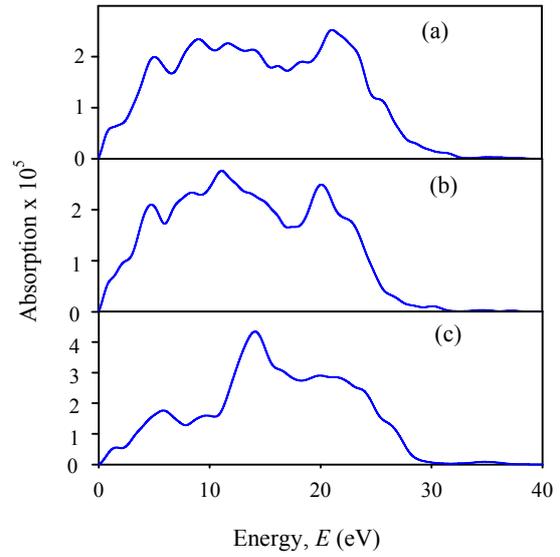

**Fig. 6.** Absorption coefficient of (a) ZnNNi$_3$, (b) SnNNi$_3$, and (c) CuNNi$_3$.

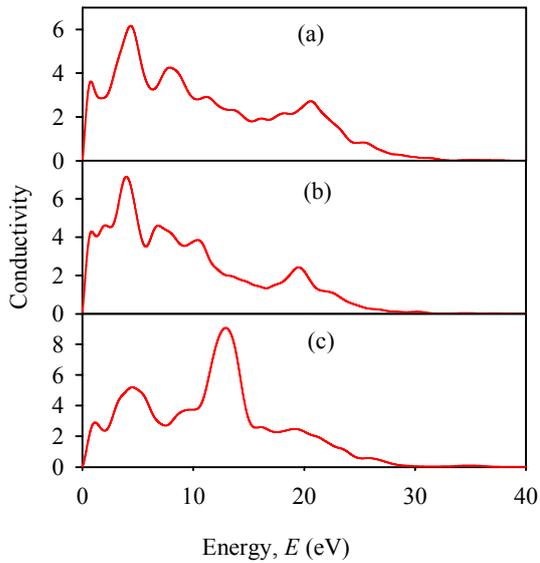

**Fig. 7.** Conductivity of (a) ZnNNi$_3$, (b) SnNNi$_3$, and (c) CuNNi$_3$.

We also observed that the calculated absorption coefficient (Fig. 6) and real part of the optical conductivity (Fig. 7) have several maxima and minima within the energy range studied. The peak structures can be explained from our band structure results. For ZnNNi$_3$, (Fig. 6a) main absorption peaks are seen with some intermediate peaks. All the spectra rapidly decrease to zero at ~ 30 eV. The peaks in the low energy region arise due to the transition from N-2$p$/Ni-3$d$ states to Ni-3$d$ states. From Fig. 7 we see that the photoconductivity starts with zero photon energy, indicating that the materials should have no band gap. Indeed no band gap is seen in the band structure calculations (Fig. 1). Moreover, the photoconductivity and hence the electrical conductivity of a material increases as a result of absorbing photons [43].

## 4. Conclusion

We have carried out a detailed investigation of the structural, elastic, electronic and optical properties of the antiperovskite-type Ni-rich nitrides MNNi$_3$ (M = Zn, Sn, and Cu) using the first-principles calculations. The analysis shows that the two hypothetical phases are good conductors and mechanically stable like the well established ZnNNi$_3$. It is shown that CuNNi$_3$ has a smaller lattice constant and larger bulk modulus compared to the other phases of MNNi$_3$. The Poisson's ratio values obtained indicate that these compounds possess metallic-like bonding. Further the electronic band structures show metallic conductivity for all the phases. It is seen that N-2$p$ and Ni-3$d$ states dominate the total DOS at the Fermi level. The prominent features in the spectra of the optical parameter have also been discussed. The large reflectivity in the low-energy region (with moderately good reflectivity up to 20~25 eV) indicates suitability of the compounds for use in the solar cell to remove solar heating. It is expected that our calculations should motivate experimental study on the two hypothetical compounds as well as experiment on optical properties of ZnNNi$_3$.